\documentstyle[prl,aps]{revtex}
\twocolumn

\begin{document}
\author{J.Q. Shen\footnote{E-mail address: jqshen@coer.zju.edu.cn}}
\address{Zhejiang Institute of Modern Physics and Department of Physics,
Zhejiang University, Hangzhou 310027, P.R. China}
\date{\today }
\title{The purely gravitational generalization of spin-rotation couplings}
\maketitle

\begin{abstract}
The nature of Mashhoon's spin-rotation coupling is the interaction
between a particle spin (gravitomagnetic moment) and a
gravitomagnetic field. Here we will consider the coupling of
graviton spin to the weak gravitomagnetic fields by analyzing the
Lagrangian density of weak gravitational field, and hence study
the purely gravitational generalization of Mashhoon's
spin-rotation couplings.

 PACS number(s): 04.20.Fy, 04.25.Nx, 04.20.Cv
\end{abstract}

With the development of laser technology and its application to
the gravitational interferometry experiments\cite{Ahmedov}, some
weak gravitational effects\cite{Werner,Atwood,Anandan,Dresden}
associated with gravitomagnetic fields has become increasingly
important and therefore captured many authors both theoretically
and experimentally. During the past 20 years, neutron
interferometry was developed with increasing accuracy. By using
these technologies, Werner {\it et al.} investigated the neutron
analog of the Foucault-Michelson-Gale effect in 1979\cite{Werner}
and Atwood {\it et al.} found the neutron Sagnac effect in
1984\cite{Atwood}. Aharonov, Carmi {\it et
al.}\cite{Anandan,Dresden,Carmi} proposed a gravitational analog
to the Aharonov-Bohm effect\cite{Bohm}, which is a geometric
effect of vector potential of gravity: specifically, in a rotating
frame the matter wave propagating along a closed path will acquire
a nonintegral phase factor (geometric phase factor). This
phenomenon has now been called the Aharonov-Carmi effect, or the
gravitational Aharonov-Bohm effect. Overhauser,
Colella\cite{Overhauser}, Werner and Standenmann {\it et al.}\cite
{Werner} have proved the existence of the Aharonov-Carmi effect by
means of the neutron-gravity interferometry experiment. Note that
here the Aharonov-Carmi effect results from the interaction
between the momentum of a moving particle and the rotating frame.
Even though the interaction of a spinning particle such as neutron
with the rotating frame has the same origin of Aharonov-Carmi
effect, {\it i.e.}, both arise from the presence of the inertial
force, the Aharonov-Carmi effect mentioned above does not contain
the spin-rotation coupling\cite{Mashhoon}.

It has become feasible to use polarized neutrons in the
interferometer experiments\cite{Hehl}.  Since a particle with an
intrinsic spin possesses a gravitomagnetic moment, Mashhoon
considered the interaction of the particle spin with the rotation
of a noninertial reference frame, which was referred to as the
{\it spin-rotation coupling}\cite{Mashhoon}. Recently, Mashhoon
analyzed the Doppler effect of wavelight in a rotating frame with
respect to the fixing frame\cite{Mashhoon1,Mashhoon2} and derived
the photon spin-rotation coupling effect, and we considered the
coordinate transformation (from the fixing frame to the rotating
frame) of gravitomagnetic potential $g_{0\varphi}$ of Kerr metric
and then obtained the Hamiltonian of neutron spin-rotation
coupling\cite{Shen}. A straightforward and unified way within the
framework of special relativity to derive the inertial effects
(including spin-rotation coupling, Bonse-Wroblewski and
Page-Werner effects) of a Dirac particle was proposed by Hehl {\it
et al.}, where they put the special-relativistic Dirac equation
into a noninertial reference frame by standard methods\cite{Hehl}.
Mashhoon's spin-rotation coupling has some interesting
applications, {\it e.g.}, the spin-rotation coupling experienced
by valency electrons in rapidly rotating C$_{60}$ molecules (with
the rotating frequency ranges from $10^{10}$ to $10^{12}$ rad/s in
the orientationally disordered phase) can be applied to the
investigations of photoelectron spectroscopy (and hence to the
molecular dynamics of C$_{60}$ solid)\cite{He}.

Basically, the spin-rotation coupling considered above is just one
of the gravitomagnetic effects, for the rotating frequency of the
noninertial frame can be viewed as a gravitomagnetic field (or a
piece of the gravitomagnetic field)\cite{Shenarxiv} according to
the principle of equivalence. To the best of our knowledge, the
spin-rotation coupling of photon, electron and neutron has been
taken into account in the literature\cite{Mashhoon2,Shen,He}.
However, the gravitational coupling of graviton spin to the
gravitomagnetic fields, which may also be of physical interest,
has so far never yet been considered. Here we will extend
Mashhoon's spin-rotation coupling to a purely gravitational case,
where the graviton spin will be coupled to the gravitomagnetic
fields. It will be shown that by analyzing the third-order terms
in Lagrangian density of weak gravitational field, one can obtain
the expression of the Lagrangian density for the interaction
between the graviton spin and gravitomagnetic fields.

In this report, we deal with the weak gravitational fields only,
which is described by the linearized gravity theory. One speaks of
a linearized theory when the metric deviates only slightly from
that of flat space. A weak gravitational field (in which spacetime
is nearly flat) is defined as a manifold on which coordinates
exist, where the metric has components
$g_{\alpha\beta}=\eta_{\alpha\beta}+h_{\alpha\beta}$ with
$|h_{\alpha\beta}|\ll 1$. Such coordinates are called nearly
Lorentz coordinates, where the indices of tensors are raised and
lowered with the flat-Minkowski metric. First, we consider the
expression for the spin of weak gravitational fields in the
linearized gravity theory. In the nearly Lorentz coordinate
system, to the first order in $h_{\mu\nu}$, the Lagrangian density
of the gravitational field takes the form
\begin{equation}
{\mathcal
L}=-2(h^{\beta\nu,\mu}h_{\mu\nu,\beta}-\frac{1}{2}h^{\beta\nu,\mu}h_{\beta\nu,\mu}-\frac{1}{2}h^{\beta\nu}_{\
\ \ ,\nu}h^{\mu}_{\ \ \mu,\beta}).   \label{lagrangian}
\end{equation}
According to the canonical procedure, the canonical momentum of
the linearized gravitational field is
\begin{equation}
\pi^{\mu\nu}=\frac{\partial {\mathcal L}}{\partial
\dot{h}_{\mu\nu}}=-4(h^{0\nu,\mu}-\frac{1}{2}h^{\mu\nu,0})+\eta^{\mu\nu}h^{0\lambda}_{\
\ \ ,\lambda}+\eta^{\nu 0}h^{,\mu},    \label{pi}
\end{equation}
where dot denotes the derivative with respect to time, and
$h=\eta^{\mu\nu}h_{\mu\nu}$. In accordance with the Noether
theorem, the spin of a field that is characterized by the
Lagrangian density (\ref{lagrangian}) is of the form $
S^{\theta\tau}=\int {\rm
d}^{3}x(\pi^{\mu\nu}\sum^{\theta\tau}_{\nu\eta}h^{\eta}_{\ \ \mu})
$, where $
\sum^{\theta\tau}_{\nu\eta}=\delta^{\theta}_{\nu}\delta^{\tau}_{\eta}-\delta^{\theta}_{\eta}\delta^{\tau}_{\nu}$.
It is well known that in the linearized gravity theory, there are
two fundamental types of coordinate transformations, which take
one nearly Lorentz coordinate system into another. The two
coordinate transformations are the background Lorentz
transformation, where the background metric takes the form of
simple diagonal $(+1,-1,-1,-1)$, and the gauge transformation (for
$h_{\mu\nu}$). To the first order in small quantities, such a
gauge transformation for $h_{\mu\nu}$ is
\begin{equation}
h'_{\alpha\beta}=h_{\alpha\beta}-\zeta_{\alpha,\beta}-\zeta_{\beta,\alpha},
\end{equation}
where $\zeta_{\alpha}$ is an infinitesimal variation in the
coordinate transformation, {\it i.e.},
$x'_{\alpha}=x_{\alpha}+\zeta_{\alpha}(x_{\beta})$. It is readily
verified that by using the appropriate gauge
conditions\cite{Shenarxiv}, the integrand (the density of spin,
$s^{\theta\tau}$, of gravitational field) in $S^{\theta\tau}$ can
be rewritten as\cite{Shenarxiv}
\begin{equation}
s^{\theta\tau}=-2\left(h^{0\tau}h^{0\theta,0}-h^{0\theta
}h^{0\tau,0}\right), \label{finalspin}
\end{equation}
where $h^{0\theta}$ ($\theta=1,2,3$) can be regarded as the
gravitomagnetic vector potentials.

Since in a rotating frame of reference, the expression
$\partial_{m}h_{0n}-\partial_{n}h_{0m}$ for the gravitomagnetic
field is such a quantity, the magnitude of which is proportional
to the angular frequency, $\omega$, of the rotating
frame\cite{Shenarxiv}. Thus, it is believed that the expression
associated with
$(h^{0m}\dot{h}^{0n}-h^{0n}\dot{h}^{0m})(\partial_{m}h_{0n}-\partial_{n}h_{0m})$
in the gravitational Lagrangian density (containing the nonlinear
terms) can be viewed as an interaction term of the generalized
Mashhoon's spin-rotation coupling, {\it i.e.}, the interaction of
the graviton spin with the gravitomagnetic fields. Here the
repeated indices implies the summation carried out over $1, 2, 3$.

Consider the Christoffel symbol of the weak gravitational field,
the first- and second- order terms of which may be expressed as
follows\cite{Shao1,Shao2}
\begin{eqnarray}
\Gamma^{\alpha}_{\ \beta
\gamma}&=&-\frac{1}{2}K(h^{\alpha}_{\beta,\gamma}+h^{\alpha}_{\gamma,\beta}-h^{,\alpha}_{\beta\gamma}-\frac{1}{2}\delta^{\alpha}_{\beta}h^{\lambda}_{\lambda,\gamma}-\frac{1}{2}\delta^{\alpha}_{\gamma}h^{\lambda}_{\lambda,\beta}    \nonumber \\
&+&\frac{1}{2}\eta_{\beta\gamma}h^{\lambda,\alpha}_{\lambda})
+\frac{1}{2}K^{2}(h_{\beta\lambda}h^{\lambda\alpha}_{,\gamma}+h_{\gamma\lambda}h^{\lambda\alpha}_{,\beta}+h^{\alpha\lambda}h_{\beta\gamma,\lambda}      \nonumber \\
&-&
h_{\beta\lambda}h^{\lambda,\alpha}_{\gamma}-h_{\gamma\lambda}h^{\lambda,\alpha}_{\beta})
+\frac{1}{2}K^{2}(-\frac{1}{2}\delta^{\alpha}_{\beta}h_{\lambda\tau}h^{\lambda\tau}_{,\gamma}     \nonumber \\
&-&\frac{1}{2}\delta^{\alpha}_{\gamma}h_{\lambda\tau}h^{\lambda\tau}_{,\beta}-\frac{1}{2}\eta_{\beta\gamma}h^{\alpha\lambda}h^{\tau}_{\tau,\lambda}+\frac{1}{2}h_{\beta\gamma}h^{\lambda,\alpha}_{\lambda}    \nonumber \\
&+&\frac{1}{2}\eta_{\beta\gamma}h_{\lambda\tau}h^{\lambda\tau,\alpha})+{\mathcal
O}(h^{3}),        \label{eqeq}
\end{eqnarray}
where $K$ and $h^{\mu\nu}$ are so defined that
$\sqrt{-g}g^{\mu\nu}=\eta^{\mu\nu}+Kh^{\mu\nu}$ is satisfied.
Because of $g^{\mu\nu}R_{\mu\nu}$ containing
$g^{0m}R_{0m}+g^{m0}R_{m0}$ ({\it i.e.}, $2g^{0m}R_{0m}$), we will
analyze only the $0m$ component of the Ricci tensor $R_{\mu\nu}$
in the following. Note that the expression for $R_{\mu\nu}$ is
written as $R_{\mu\nu}=\Gamma^{\rho}_{\
\mu\rho,\nu}-\Gamma^{\rho}_{\ \mu\nu,\rho}-\Gamma^{\sigma}_{\
\mu\nu}\Gamma^{\rho}_{\ \sigma\rho}+\Gamma^{\rho}_{\
\sigma\nu}\Gamma^{\sigma}_{\ \mu\rho}$. In what follows, for
brevity, the above generalized interaction will be referred to as
the {\it graviton S-G coupling}. Now we will extract the
expression $\sim
\dot{g}^{0n}(\partial_{m}g_{0n}-\partial_{n}g_{0m})$ from the
Ricci tensor $R_{\mu\nu}$.

\subsection{$\Gamma^{\rho}_{\ \mu\rho,\nu}-\Gamma^{\rho}_{\ \mu\nu,\rho}$}
It follows from the expression (\ref{eqeq}) that the second-order
terms (with the coefficient being $\frac{1}{2}K^{2}$), which will
probably contribute to the graviton S-G coupling, in the
derivative $\Gamma^{\rho}_{\ \mu\rho,\nu}$, are given as follows $
h_{\mu\lambda,\nu}h^{\lambda\rho}_{,\rho}+h_{\rho\lambda,\nu}h^{\lambda\rho}_{,\mu}+h^{\rho\lambda}_{,\nu}h_{\mu\rho,\lambda}-h_{\mu\lambda,\nu}h^{\lambda,\rho}_{\rho}-h_{\rho\lambda,\nu}h^{\lambda,\rho}_{\mu}-\frac{5}{2}h_{\lambda\tau,\nu}h^{\lambda\tau}_{,\mu}-\frac{1}{2}h^{\lambda}_{\mu,\nu}h^{\tau}_{\tau,\lambda}+\frac{1}{2}h_{\mu\rho,\nu}h^{\lambda,\rho}_{\lambda}+\frac{1}
{2}h_{\lambda\tau,\nu}h^{\lambda\tau}_{,\mu}$. But the detailed
analysis shows that the terms that truly gives contribution to the
graviton S-G coupling are
$-h_{\rho\lambda,\nu}h^{\lambda\rho}_{,\mu}$ and
$h_{\rho\lambda,\nu}\left(h_{\mu}^{\rho,\lambda}-h_{\mu}^{\lambda,\rho}\right)$
only, the latter of which, however, vanishes. So, when taking into
account the Lagrangian density $\sqrt{-g}g^{\mu\nu}R_{\mu\nu}$ of
gravitational field, it is clearly seen that
$-\sqrt{-g}g^{\mu\nu}h_{\rho\lambda,\nu}h^{\lambda\rho}_{,\mu}$
contains the expression
$\sqrt{-g}g^{0m}\left[-h_{\rho\lambda,0}h^{\lambda\rho}_{,m}-h_{\rho\lambda,m}h^{\lambda\rho}_{,0}\right]$
({\it i.e.},
$-2\sqrt{-g}g^{0m}h_{\rho\lambda,0}h^{\lambda\rho}_{,m}$), which
contains
\begin{eqnarray}
& &
\sqrt{-g}g^{0m}\left[-2h_{0n,0}h^{n0}_{,m}-2h_{n0,0}h^{0n}_{,m}\right]
\nonumber  \\
&=&-4\sqrt{-g}g^{0m}\dot{g}_{n}h^{n0}_{,m}=4\sqrt{-g}g^{0m}\dot{g}_{n}\partial_{m}g_{n}.
\label{eq40}
\end{eqnarray}
Here $h^{0n}=g^{n}=-g_{n}$. The flat Minkowski metric
$\eta_{\mu\nu}={\rm diag}[+1, -1,-1,-1]$. Clearly, the relation
between the metric $g^{0m}$ (with $m=1,2,3$) and the
gravitomagnetic vector potential $g^{m}$ is that
$\sqrt{-g}g^{0m}=Kh^{0m}=Kg^{m}$.

It is readily verified that the second-order terms in
$\Gamma^{\rho}_{\ \mu\nu,\rho}$ which may have effect on the
graviton S-G coupling are
$\left(h_{\mu\lambda,\rho}h^{\lambda\rho}_{,\nu}+h_{\nu\lambda,\rho}h^{\lambda\rho}_{,\mu}\right)+h^{\rho\lambda}_{,\rho}h_{\mu\nu,\lambda}-\left(h_{\mu\lambda,\rho}h^{\lambda,\rho}_{\nu}+h_{\nu\lambda,\rho}h^{\lambda,\rho}_{\mu}\right)-\frac{1}{2}\left(h_{\lambda\tau,\mu}h^{\lambda\tau}_{,\nu}+h_{\lambda\tau,\nu}h^{\lambda\tau}_{,\mu}\right)
-\frac{1}{2}\eta_{\mu\nu}h^{\rho\lambda}_{,\nu}h^{\tau}_{\tau,\lambda}+\frac{1}{2}h_{\mu\nu,\rho}h^{\lambda,\rho}_{\lambda}+\frac{1}{2}\eta_{\mu\nu}h_{\lambda\tau,\rho}h^{\lambda\tau,\rho}
$. Further analysis demonstrates that the terms which truly give
contribution to the graviton S-G coupling are the following four
terms:

(i)
$\sqrt{-g}g^{\mu\nu}\left(h_{\mu\lambda,\rho}h^{\lambda\rho}_{,\nu}+h_{\nu\lambda,\rho}h^{\lambda\rho}_{,\mu}\right)$
contains
$\sqrt{-g}g^{0m}\left(h_{0\lambda,\rho}h^{\lambda\rho}_{,m}+h_{m\lambda,\rho}h^{\lambda\rho}_{,0}+h_{0\lambda,\rho}h^{\lambda\rho}_{,m}
+h_{m\lambda,\rho}h^{\lambda\rho}_{,0}\right)$, {\it i.e.},
$2\sqrt{-g}g^{0m}\left(h_{0\lambda,\rho}h^{\lambda\rho}_{,m}+h_{m\lambda,\rho}h^{\lambda\rho}_{,0}\right)$,
which includes the following terms
\begin{eqnarray}
& &
2\sqrt{-g}g^{0m}\left(h_{0n,0}h^{n0}_{,m}+h_{m0,n}h^{0n}_{,0}\right)
\nonumber  \\
&=&-2\sqrt{-g}g^{0m}\dot{g}_{n}\left(\partial_{m}g_{n}+\partial_{n}g_{m}\right).
\label{eq1}
\end{eqnarray}

(ii)
$-\sqrt{-g}g^{\mu\nu}\left(h_{\mu\lambda,\rho}h^{\lambda,\rho}_{\nu}+h_{\nu\lambda,\rho}h^{\lambda,\rho}_{\mu}\right)$
contains $
-\sqrt{-g}g^{0m}\left(h_{0\lambda,\rho}h^{\lambda,\rho}_{m}+h_{m\lambda,\rho}h^{\lambda,\rho}_{0}
+h_{0\lambda,\rho}h^{\lambda,\rho}_{m}+h_{m\lambda,\rho}h^{\lambda,\rho}_{0}\right)
$, {\it i.e.},
$-2\sqrt{-g}g^{0m}\left(h_{0\lambda,\rho}h^{\lambda,\rho}_{m}+h_{m\lambda,\rho}h^{\lambda,\rho}_{0}\right)
$, which will give no contribution to the graviton S-G coupling.
So here we will not further consider it.

(iii)
$\sqrt{-g}g^{\mu\nu}h^{\rho\lambda}_{,\rho}h_{\mu\nu,\lambda}$
contains the components $
\sqrt{-g}g^{0m}\left(h^{\rho\lambda}_{,\rho}h_{0m,\lambda}+h^{\rho\lambda}_{,\rho}h_{m0,\lambda}\right)$,
{\it i.e.}, $2\sqrt{-g}g^{0m}h^{\rho\lambda}_{,\rho}h_{0m,\lambda}
$, which includes
\begin{equation}
2\sqrt{-g}g^{0m}h^{0n}_{,0}h_{0m,n}=-2\sqrt{-g}g^{0m}\dot{g}_{n}\partial_{n}g_{m}.
\label{eq3}
\end{equation}

(iv)
$-\frac{1}{2}\sqrt{-g}g^{\mu\nu}\left(h_{\lambda\tau,\mu}h^{\lambda\tau}_{,\nu}+h_{\lambda\tau,\nu}h^{\lambda\tau}_{,\mu}\right)$
includes
$\sqrt{-g}g^{0m}\left[-h_{\lambda\tau,0}h^{\lambda\tau}_{,m}-h_{\lambda\tau,m}h^{\lambda\tau}_{,0}\right]$,
 which can be rewritten as
$-2\sqrt{-g}g^{0m}h_{\lambda\tau,0}h^{\lambda\tau}_{,m} $, and
contains
\begin{eqnarray}
& &
\sqrt{-g}g^{0m}\left[-2h_{0n,0}h^{0n}_{,m}-2h_{n0,0}h^{n0}_{,m}\right]
\nonumber  \\
&=&-4\sqrt{-g}g^{0m}h_{0n,0}h^{0n}_{,m}=4\sqrt{-g}g^{0m}\dot{g}_{n}\partial_{m}g_{n}.
\label{eq4}
\end{eqnarray}

Thus it follows from Eq.(\ref{eq1})-(\ref{eq4}) that the terms in
$\sqrt{-g}g^{\mu\nu}\Gamma^{\rho}_{\ \mu\nu,\rho}$, which
contribute to the graviton S-G coupling, are given as follows
\begin{eqnarray}
&&
\sqrt{-g}g^{\mu\nu}\left[-2\dot{g}_{n}\left(\partial_{m}g_{n}+\partial_{n}g_{m}\right)-2\dot{g}_{n}\partial_{n}g_{m}+4\dot{g}_{n}\partial_{m}g_{n}\right]
\nonumber   \\
&=&
\sqrt{-g}g^{\mu\nu}\left[2\dot{g}_{n}\partial_{m}g_{n}-4\dot{g}_{n}\partial_{n}g_{m}\right].
\label{eq140}
\end{eqnarray}

Hence it follows from (\ref{eq40}) and (\ref{eq140}) that the
terms in $\sqrt{-g}g^{\mu\nu}\left(\Gamma^{\rho}_{\
\mu\rho,\nu}-\Gamma^{\rho}_{\ \mu\nu,\rho}\right)$ which
contribute to the graviton S-G coupling are written in the form
\begin{eqnarray}
& &
\sqrt{-g}g^{0m}\left[4\dot{g}_{n}\partial_{m}g_{n}-\left(2\dot{g}_{n}\partial_{m}g_{n}-4\dot{g}_{n}\partial_{n}g_{m}\right)\right]
\nonumber   \\
&=&2\sqrt{-g}g^{0m}\left(\dot{g}_{n}\partial_{m}g_{n}+2\dot{g}_{n}\partial_{n}g_{m}\right).
\label{15}
\end{eqnarray}

\subsection{$-\Gamma^{\sigma}_{\ \mu\nu}\Gamma^{\rho}_{\
\sigma\rho}$}

The first-order terms (proportional to $-\frac{1}{2}K$) in
$\Gamma^{\sigma}_{\ \mu\nu}$ is of the form $ \Gamma^{\sigma}_{\
\mu\nu}\left(\propto
-\frac{1}{2}K\right)=h^{\sigma}_{\mu,\nu}+h^{\sigma}_{\nu,\mu}-h^{,\sigma}_{\mu\nu}
-\frac{1}{2}\delta^{\sigma}_{\mu}h^{\lambda}_{\lambda,\nu}-\frac{1}{2}\delta^{\sigma}_{\nu}h^{\lambda}_{\lambda,\mu}
+\frac{1}{2}\eta_{\mu\nu}h^{\lambda,\sigma}_{\lambda}$, which
includes the {\it valuable} terms
$-\frac{1}{2}K\left(h^{\sigma}_{\mu,\nu}+h^{\sigma}_{\nu,\mu}-h^{,\sigma}_{\mu\nu}\right)$.
In the meanwhile, the terms proportional to $-\frac{1}{2}K$ in
$\Gamma^{\rho}_{\ \sigma\rho}$ take the form $ \Gamma^{\rho}_{\
\sigma\rho}\left(\propto
-\frac{1}{2}K\right)=-\frac{1}{2}K\left(h^{\rho}_{\sigma,\rho}+h^{\rho}_{\rho,\sigma}
-h^{,\rho}_{\sigma\rho}+...\right)=-\frac{1}{2}Kh^{\rho}_{\rho,\sigma}
$.

It is readily verified that the contribution of
$-\Gamma^{\sigma}_{\ \mu\nu}\Gamma^{\rho}_{\ \sigma\rho}$ to the
graviton S-G coupling is vanishing. So, here we will not consider
it further.

\subsection{$\Gamma^{\rho}_{\ \sigma\nu}\Gamma^{\sigma}_{\
\mu\rho}$}

It is apparently seen that $\sqrt{-g}g^{\mu\nu}\Gamma^{\rho}_{\
\sigma\nu}\Gamma^{\sigma}_{\ \mu\rho}$ includes the following
terms $ \sqrt{-g}g^{0m}\left(\Gamma^{\rho}_{\
\sigma0}\Gamma^{\sigma}_{\ m\rho}+\Gamma^{\rho}_{\ \sigma
m}\Gamma^{\sigma}_{\
0\rho}\right)=\sqrt{-g}g^{0m}\left(\Gamma^{\rho}_{\
\sigma0}\Gamma^{\sigma}_{\ m\rho}+\Gamma^{\rho}_{\
0\sigma}\Gamma^{\sigma}_{\ \rho
m}\right)=2\sqrt{-g}g^{0m}\Gamma^{\rho}_{\ \sigma
0}\Gamma^{\sigma}_{\ m\rho}$, where $ \Gamma^{\rho}_{\ \sigma
0}\left(\propto
-\frac{1}{2}K\right)=h^{\rho}_{\sigma,0}+h^{\rho}_{0,\sigma}-h^{,\rho}_{\sigma
0}$ and $ \Gamma^{\sigma}_{\ m\rho}\left(\propto
-\frac{1}{2}K\right)=h^{\sigma}_{m,\rho}+h^{\sigma}_{\rho,
m}-h^{,\sigma}_{m\rho}$. Thus, $\Gamma^{\rho}_{\ \sigma
0}\Gamma^{\sigma}_{\ m\rho}$ contains $
\left(h^{\rho}_{\sigma,0}+h^{\rho}_{0,\sigma}-h^{,\rho}_{\sigma
0}\right)\left(h^{\sigma}_{m,\rho}+h^{\sigma}_{\rho,
m}-h^{,\sigma}_{m\rho}\right)$, which is rewritten as
$\left[h^{\rho}_{\sigma,0}\left(h^{\sigma}_{\rho,
m}-h^{,\sigma}_{m\rho}\right)\right]+h^{\rho}_{\sigma,0}h^{\sigma}_{m,\rho}
+\left(h^{\rho}_{0,\sigma}-h^{,\rho}_{\sigma
0}\right)\left(h^{\sigma}_{m,\rho}+h^{\sigma}_{\rho,
m}-h^{,\sigma}_{m\rho}\right)$. In the following discussions, for
convenience, we classify the terms in
$\left(h^{\rho}_{\sigma,0}+h^{\rho}_{0,\sigma}-h^{,\rho}_{\sigma
0}\right)\left(h^{\sigma}_{m,\rho}+h^{\sigma}_{\rho,
m}-h^{,\sigma}_{m\rho}\right)$ into three categories:

(i) $h^{\rho}_{\sigma,0}\left(h^{\sigma}_{\rho,
m}-h^{,\sigma}_{m\rho}\right)$ contains
\begin{equation}
h^{0}_{n,0}\left(h^{n}_{0,
m}-h^{,n}_{m0}\right)=-\dot{g}_{n}\left(\partial_{m}g_{n}-\partial_{n}g_{m}\right).
\label{eq22}
\end{equation}

(ii) $h^{\rho}_{\sigma,0}h^{\sigma}_{m,\rho}$ contains
\begin{equation}
h^{n}_{0,0}h^{0}_{m,n}=-\dot{g}_{n}\partial_{n}g_{m}. \label{eq23}
\end{equation}

(iii) $\left(h^{\rho}_{0,\sigma}-h^{,\rho}_{\sigma
0}\right)\left(h^{\sigma}_{m,\rho}+h^{\sigma}_{\rho,
m}-h^{,\sigma}_{m\rho}\right)$ contains the following six terms
\begin{eqnarray}
& & h^{\rho}_{0,\sigma}h^{\sigma}_{m,\rho}=
h^{n}_{0,0}h^{0}_{m,n}+h^{0}_{0,0}h^{0}_{m,0}\Rightarrow -\dot{g}_{n}\partial_{n}g_{m},                 \nonumber \\
& & h^{\rho}_{0,\sigma}h^{\sigma}_{\rho, m}=h^{n}_{0,0}h^{0}_{n,
m}+h^{0}_{0,0}h^{0}_{0, m}\Rightarrow
-\dot{g}_{n}\partial_{m}g_{n},          \nonumber \\
& & -h^{\rho}_{0,\sigma}h^{,\sigma}_{m\rho} \ ({\rm
giving \ no \ contribution \ to \ S-G \ coupling}),          \nonumber \\
& & -h^{,\rho}_{\sigma 0}h^{\sigma}_{m,\rho} \ ({\rm
giving \ no \ contribution \ to \ S-G \ coupling}),          \nonumber \\
& & -h^{,\rho}_{\sigma 0}h^{\sigma}_{\rho,m}=-h^{,0}_{n 0}h^{n}_{0,m}-h^{,0}_{0 0}h^{0}_{0,m}
\Rightarrow \dot{g}_{n}\partial_{m}g_{n},     \nonumber \\
& & \left(-h^{,\rho}_{\sigma
0}\right)\left(-h^{,\sigma}_{m\rho}\right)=h_{n0}^{,0}h_{m0}^{,n}+h_{00}^{,0}h_{m0}^{,0}\Rightarrow
-\dot{g}_{n}\partial_{n}g_{m}.           \label{eq24}
\end{eqnarray}
So, the terms in $\left(h^{\rho}_{0,\sigma}-h^{,\rho}_{\sigma
0}\right)\left(h^{\sigma}_{m,\rho}+h^{\sigma}_{\rho,
m}-h^{,\sigma}_{m\rho}\right)$, which will have effect on the
graviton S-G coupling, are
\begin{equation}
-2\dot{g}_{n}\partial_{n}g_{m}   \label{eq25}
\end{equation}
only.

Thus it follows from (\ref{eq22}), (\ref{eq23}) and (\ref{eq25})
that the total terms in $\sqrt{-g}g^{\mu\nu}\Gamma^{\rho}_{\
\sigma\nu}\Gamma^{\sigma}_{\ \mu\rho}$ which will give
contribution to the graviton S-G coupling are expressed as follows
\begin{eqnarray}
& &
2\sqrt{-g}g^{0m}\left[-\dot{g}_{n}\left(\partial_{m}g_{n}-\partial_{n}g_{m}\right)-\dot{g}_{n}\partial_{n}g_{m}-2\dot{g}_{n}\partial_{n}g_{m}\right]
\nonumber  \\
&=&2\sqrt{-g}g^{0m}\left[-\dot{g}_{n}\left(\partial_{m}g_{n}-\partial_{n}g_{m}\right)-3\dot{g}_{n}\partial_{n}g_{m}\right].
\label{26}
\end{eqnarray}

\subsection{The final result}
Hence, according to the expressions (\ref{15}) and (\ref{26}), one
can finally obtain the total contribution to the graviton S-G
coupling as $\frac{1}{2}K^{2}\cdot
2\sqrt{-g}g^{0m}\left(\dot{g}_{n}\partial_{m}g_{n}+2\dot{g}_{n}\partial_{n}g_{m}\right)+\frac{1}{4}K^{2}\cdot
2\sqrt{-g}g^{0m}\left[-\dot{g}_{n}\left(\partial_{m}g_{n}-\partial_{n}g_{m}\right)-3\dot{g}_{n}\partial_{n}g_{m}\right]$,
{\it i.e.},
$\frac{1}{2}K^{2}\sqrt{-g}g^{0m}\left[-2\dot{g}_{n}\left(\partial_{m}g_{n}-\partial_{n}g_{m}\right)+3\dot{g}_{n}\partial_{m}g_{n}\right]$.
Thus one can arrive at
\begin{eqnarray}
{\mathcal
L}_{s-g}&=&-K^{2}\sqrt{-g}g^{0m}\dot{g}_{n}\left(\partial_{m}g_{n}-\partial_{n}g_{m}\right)       \nonumber   \\
& =&
\frac{1}{2}K^{3}\left(g_{m}\dot{g}_{n}-g_{n}\dot{g}_{m}\right)\left(\partial_{m}g_{n}-\partial_{n}g_{m}\right),
\label{eqfinal}
\end{eqnarray}
where the relation $\sqrt{-g}g^{0m}=Kh^{0m}=Kg^{m}=-Kg_{m}$ has
been inserted, and the summation for the repeated indices is
carried out over the values $1, 2, 3$. Thus we present the
Lagrangian density that describes the coupling of graviton spin
(spinning moment) to gravitomagnetic fields, the generalized
version of Mashhoon's spin-rotation couplings in the purely
gravitational case. It may be believed that such a graviton S-G
coupling deserves further detailed investigation, for the coupling
of graviton spin to gravitomagnetic fields may provide us with a
deep insight into Mashhoon's spin-rotation
couplings\cite{Mashhoon2} (and hence into the inertial effects of
spinning particles)\cite{Hehl}.

In addition, since the purely gravitational generalization of
Mashhoon's spin-rotation coupling, {\it i.e.}, the interaction of
the graviton spin with the gravitomagnetic fields is actually a
self-interaction of the spacetime (gravitational fields), here we
will propose a definite prediction about such a purely
gravitational interaction, which will arise in a noninertial frame
of reference itself: specifically, a rotating frame that
experiences a fluctuation of its rotational frequency will undergo
such a weak self-interaction. In other words, such a
self-interaction of the rotating frame, which can also be called
the self-interaction of the spacetime of the rotating frame, is
just a noninertial generalization of the interaction of the
graviton spin with the gravitomagnetic fields. This can be
understood as follows: in a rotating frame, the
$\varphi$-component of the gravitomagnetic vector potential is
$g_{\varphi }\simeq (2\omega r/c)\sin \theta$\cite{Shenarxiv}. It
follows that if either the direction or the magnitude of the
angular velocity of the rotating frame changes, then the time
derivative of the gravitomagnetic vector potentials is
nonvanishing, {\it e.g.}, $ |\dot{g}_{\varphi }|\simeq
|(2\dot{\omega} r/c)\sin \theta|\neq 0 $, and according to
(\ref{eqfinal}), such a rotating frame will thus be subjected to a
self-interaction, the nature of which is just a nonlinear
interaction of the spacetime of the noninertial frame of reference
itself. We suggest that the precessional motion of the rapidly
rotating C$_{60}$ molecules may probably provide us with an ideal
way to test such a self-interaction of the noninertial frame. In
the high-temperature phase ({\it i.e.}, the orientationally
disordered phase), the rate of change of the angular velocity
$\vec{\omega}$ of C$_{60}$ acted upon by the noncentral
intermolecular forces is $10^{21}\sim 10^{23}{\rm s}^{-2}$
\cite{Shenarxiv}. So, the linear acceleration of valency electrons
on the C$_{60}$ molecular surface due to the above fluctuation in
$\vec{\omega}$ is about $ 10^{12}\sim 10^{14}$ m$/{\rm s}^{2}$,
which is the same order of magnitude of the inertial centrifugal
acceleration due to the rapid rotation of C$_{60}$ (in the
orientationally disordered phase, the rotational frequency of
C$_{60}$ molecules is about $10^{11}$ rad/s\cite{He}). It is known
that the weak gravitational effects, including the Aharonov-Carmi
effect\cite{Carmi}, neutron Sagnac effect\cite{Post,Horne},
spin-rotation coupling\cite{Mashhoon,Mashhoon1,Mashhoon2} and so
on are in fact the inertial effects of the rotating frame. Most of
these inertial effects may have influence on the frequency
(energy) shift in atoms, molecules and light
wave\cite{Mashhoon2,He,Post}. Likewise, since both the fluctuation
in the angular velocity ({\it i.e.}, the nutational and
precessional motions) of C$_{60}$ molecules and the change in the
gravitomagnetic vector potential (measured by the observer fixed
in the C$_{60}$ rotating frame) is very great, we think that the
above-mentioned self-interaction of C$_{60}$ rotating frame might
deserve investigation for the treatment of the photoelectron
spectroscopy, noncentral intermolecular potential and molecular
rotational dynamics of C$_{60}$ solid.
\\ \\
\textbf{Acknowledgements}  This project was supported partially by
the National Natural Science Foundation of China under Project No.
$90101024$.

\end{document}